# Advanced Page Rank Algorithm with Semantics, In Links, Out Links and Google Analytics

Aritra Banerjee[1], Shrey Choudhary[2]
*4th year B. Tech CSE students*
*Vit University Chennai Campus, Vandalur-Kelambakkam Road, Chennai – 600048, Tamil Nadu, India*

***Abstract-*** *In this paper, we have modified the existing page ranking mechanism as an advanced Page Rank Algorithm based on Semantics, Inlinks, outlinks and Google Analytics. We have used Semantics page ranking to rank pages according to the word searched and match it with the metadata of the website and provide a value of rank according to the highest priority. We have also used Google analytics to store the number of hits of a website in a particular variable and add the required percentage amount to the ranking procedure. The proposed algorithm is used to find more relevant information according to user's query. So, this concept is very useful to display most valuable pages on the top of the result list on the basis of user browsing behaviour, which reduce the search space to a large scale.*

## I. INTRODUCTION

PageRank is a link analysis algorithm and it assigns a numerical weighting to each element of a hyperlinked set of documents, such as the World Wide Web (WWW), with the purpose of "measuring" its relative importance within the set. The algorithm may be applied to any collection of entities with reciprocal quotations and references. The numerical weight that it assigns to any given element *E* is referred to as the *PageRank of E* and denoted by PR (E). Other factors like *Author Rank* can contribute to the importance of an entity.

## II. EXISTING WORK

**Brin and Page[1]** developed PageRank algorithm at Stanford University based on the hyper link structure. PageRank algorithm is used by the famous search engine, Google. PageRank algorithm is the most frequently used algorithm for ranking billions of web pages. During the processing of a query, Google's search algorithm combines precomputed PageRank scores with text matching scores to obtain an overall ranking score for each web page. Functioning of the Page Rank algorithm depends upon link structure of the web pages. The PageRank algorithm is based on the concepts that if a page surrounds important links towards it then the links of this page near the other page are also to be believed as imperative pages. The Page Rank imitate on the back link in deciding the rank score. Thus, a page gets hold of a high rank if the addition of the ranks of its back links is high. A simplified version of PageRank is given in Eq. 1

$$PR(u) = c \sum_{v \in B(u)} \frac{PR(v)}{N_v} \qquad (1)$$

Where u represents a web page, B (u) is the set of pages that point to u, PR (u) and PR (v) are rank scores of page u and v respectively, Nv indicates the number of outgoing links of page v, c is a factor applied for normalization. Later PageRank was customized observing that not all users follow the direct links on WWW. The modified version is given in Eq.2:

$$PR(u) = (1-d) + d \sum_{v \in B(u)} \frac{PR(v)}{N_v} \qquad (2)$$

Where d is a dampening factor that is frequently set to 0.85. d can be thought of as the prospect of users' following the direct links and (1 –d) as the page rank distribution from non- directly linked pages.

**Wenpu Xing et al [2]** discussed a new approach known as weighted page rank algorithm (WPR). This algorithm is an extension of PageRank algorithm. WPR takes into account the importance of both the inlinks and the outlinks of the pages and distributes rank scores based on the popularity of the pages. WPR performs better than the conventional PageRank algorithm in terms of returning larger number of relevant pages to a given query. According to author the more popular webpages are the more linkages that other webpages tend to have to them or are linked to by them. The proposed extended PageRank algorithm– a Weighted PageRank Algorithm– assigns larger rank values to more important (popular) pages instead of dividing the rank value of a page evenly among its outlink pages. Each outlink page gets a value proportional to its popularity (its number of inlinks and outlinks). The popularity from the number of inlinks and outlinks is recorded as *Win* (*v,u*) and *Wout*(*v,u*), respectively. *Win*(*v,u*) given in eq. (3) is the weight of *link*(*v, u*) calculated based on the number of inlinks of page *u* and the number of inlinks of all reference pages of page *v*.

$$W^{in}_{(v,u)} = \frac{I_u}{\sum_{p \in R(v)} I_p} \quad (3)$$

Where *Iu* and *Ip* represent the number of inlinks of page *u* and page *p*, respectively. *R* (*v*) denotes the reference page list of page *v*. *Wout* (*v,u*) given in eq. (4) is the weight of *link*(*v, u*) calculated based on the number of outlinks of page *u* and the number of outlinks of all reference pages of page *v*.

$$W^{out}_{(v,u)} = \frac{O_u}{\sum_{p \in R(v)} O_p} \quad (4)$$

Where *Ou* and *Op* represent the number of outlinks of page *u* and page *p*, respectively. *R* (*v*) denotes the reference page list of page *v*. Considering the importance of pages, the original PageRank formula is modified in eq. (5) as

$$PR(u) = (1-d) + d \sum_{v \in B(u)} PR(v) W^{in}_{(v,u)} W^{out}_{(v,u)} \quad (5)$$

**Gyanendra Kumar et al [3]** proposed a new algorithm in which they considered user's browsing behavior. As most of the ranking algorithms proposed are either link or content oriented in which consideration of user usage trends are not available. In this paper, a page ranking mechanism called Page Ranking based on Visits of Links(VOL) is being devised for search engines, which works on the basic ranking algorithm of Google, i.e. PageRank and takes number of visits of inbound links of web pages into account. This concept is very useful to display most valuable pages on the top of the result list on the basis of user browsing behavior, which reduce the search space to a large scale. In this paper as the author describe that in the original PageRank algorithm, the rank score of page p, is evenly divided among its outgoing links or we can say for a page, an inbound links brings rank value from base page, p. So, he proposed an improved PageRank algorithm. In this algorithm we assign more rank value to the outgoing links which is most visited by users. In this manner a page rank value is calculated based on visits of inbound links. The modified version based on VOL is given in equation (6)

$$PR(u) = (1-d) + d \sum_{v \in B(u)} \frac{L_u PR(v)}{TL(v)} \quad (6)$$

Notations are:
☐ d is a dampening factor,
☐ u represents a web page,
☐ B (u) is the set of pages that point to u,
☐ PR (u) and PR (v) are rank scores of page u and v respectively,
☐ Lu is the number of visits of link which is pointing page u from v.
☐ TL (v) denotes total number of visits of all links present on v.

**Neelam Tyagi et al [4]** proposed a new algorithm based on the Visit of Links (VOL). Each outlink page gets a value proportional to its popularity (its number of inlinks and outlinks). The popularity from the number of inlinks and outlinks is recorded as *Win* (*v,u*) and *Wout*(*v,u*), respectively. Here they suggested an improved Weighted PageRank algorithm. In this algorithm they assigned more rank value to the outgoing links which is most visited by users and received higher popularity from number of inlinks. They did not consider here the popularity of outlinks which is considered in the original algorithm. The advanced approach in the new algorithm is to determine the user's usage trends. The user's browsing behaviour can be calculated by number of hits (visits) of links. The modified version based on WPR (VOL) is given in eq. (7)

$$WPR_{vol}(u) = (1-d) + d \sum_{v \in B(u)} \frac{L_u WPR_{vol}(v) W^{in}_{(v,u)}}{TL(v)} \quad (7)$$

Notations are:
☐ d is a dampening factor,
☐ u represents a web page,
☐ B (u) is the set of pages that point to u,
☐ WPRVOL (u) and WPRVOL (v) are rank scores of page u and v respectively,
☐ Lu is the number of visits of link which is pointing page u from v.
☐ TL (v) denotes total number of visits of all links present on v.

### III. GRAPH GENERATED

**Narsingh, Deo [6]** is the book in which we learnt the usage of Graph Theory. The graph generated below is created using the Graph Tea Software.

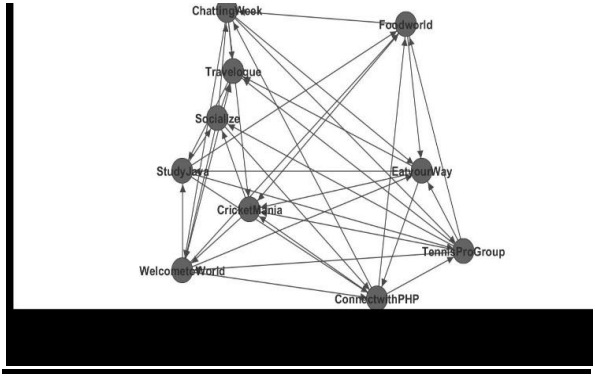

## IV. META DATA GENERATED

### *Chatting Week*

Chatting Week, Chat24*7 Chat, Private Chat, Public Chat

### *Socialize*

Socialize, Friends, Strangers, Blog, Share

### *Food World*

Food World, Pasta, Pizza, Burger, Continental, Food

### *Eat Your Way*

Eat Your Way, Paratha, Indian Food, Dosa, Parotta

### *Study Java*

Study Java, Java, JVM, Java Byte Code

### *Connect with PHP*

Connect with PHP, PHP, Backend, Cookies, Session

### *Travelogue*

Travelogue, Explore India, Travel Bangalore, Travel Hyderabad

### *Welcome to World*

Welcome to World, Explore World, Travel Paris, Travel Los Angeles

### *Cricket Mania*

Cricket Mania, IPL, World Cup, Cricket Live, Cricket Score

### *Tennis Pro Group*

Tennis Pro Group, Grand Slam, Tennis Open Live, Tennis Score, Tennis Wimbledon

## V. BINARY SEARCH ALGORITHM

**Robin Nixon [5]** is the book in which we learnt how to develop websites and how to use the languages CSS, HTML and JavaScript. We wrote the code for Binary Search using this book and also learnt about the advantage of using it.

```
function binarySearch(array, key) {

  var lo = 0,

    hi = array.length - 1,

    mid,

    element;

  while (lo <= hi) {

    mid = Math.floor((lo + hi) / 2, 10);

    element = array[mid];

    if (element < key) {

      lo = mid + 1;

    } else if (element > key) {

      hi = mid - 1;

    } else {

      return mid;

    }

  }

  return -1;

}
```

## VI. SNAPSHOT OF THE WEBPAGE

We have given a sample screenshot of the webpage which we have created using HTML and CSS mostly. Again we have used the book of **Robin Nixon [5]** in this scenario.

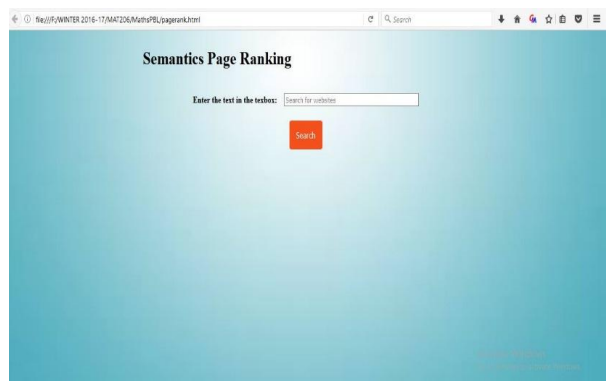

## VII. DESIGNED ALGORITHM

Let us take some websites from the given graph to calculate to its page rank:

For "FoodWorld",

No. of outlinks = 3;

No. of inlinks = 4;

Google analytics hits = 5010

For "StudyJava",

No. of outlinks = 4;

No. of inlinks = 3;

Google analytics hits = 3500

For "TennisPro",

No. of outlinks = 3;

No. of inlinks = 5;

Google analytics hits = 2000

For "Travelogue",

No. of outlinks = 4;

No. of inlinks = 4;

Google analytics hits = 2500

For "Socialize",

No. of outlinks = 3;

No. of inlinks = 6;

Google analytics hits = 7500.

Let the word searched in the textbox be **"Pasta".** As we clearly see that "Pasta" is a part of the Meta data of "FoodWorld". If the word is matched we use the counter variable 'c' and increment it to 1. If the word is not matched the counter variable 'c' is kept as 0. We take the value of dampening factor, **d = 0.3** for all calculations. We divide the weighted percentage among semantics, Google analytics and visit of links as:Semantics = 65%;Visit of Links = 25%; Google analytics = 10%. During the rank score calculation we normalise the Google analytics part only by dividing it by 1000;

**Formula for rank score:**

**Rank score = c*s + d*vl*(inlinks+outlinks) + (ga*No. of hits)/1000;**

**Calculation of Rank Score for "FoodWorld":**

Since, the word is matched with Meta data,

We put c =1;

Semantics, s = 0.65;

Visit of links, vl = 0.25;

Google analytics, ga = 0.1;

Rank Score = c*s + d*vl*(inlinks+outlinks) + (ga*6000)/1000;

$= 1*0.65 + 0.3*(0.25(4+3)) + (0.1*6000)/1000;$

$= 0.65 + 0.525 + 0.6;$

$= 1.775$

**Calculation of Rank Score for "StudyJava":**

Since, the word is not matched with Meta data,

We put c =0;

Semantics, s = 0.65;

Visit of links, vl = 0.25;

Google analytics, ga = 0.1;

Rank Score = c*s + vl*(inlinks+outlinks) + (ga*3500)/1000;

= 0*0.65 + 0.3*(0.25(4+3) )+ (0.1*3500)/1000;

= 0.00 + 0.525 + 0.35;

= 0.875

**Calculation of Rank Score for "TennisPro":**

Since, the word is not matched with Meta data,

We put c =0;

Semantics, s = 0.65;

Visit of links, vl = 0.25;

Google analytics, ga = 0.1;

Rank Score = c*s + vl*(inlinks+outlinks) + (ga*2000)/1000;

= 0*0.65 + 0.3*(0.25(3+5) )+ (0.1*2000)/1000;

= 0.00 + 0.6 + 0.2;

= 0.8

**Calculation of Rank Score for "Travelogue":**

Since, the word is not matched with Meta data,

We put c =0;

Semantics, s = 0.65;

Visit of links, vl = 0.25;

Google analytics, ga = 0.1;

Rank Score = c*s + vl*(inlinks+outlinks) + (ga*2500)/1000;

= 0*0.65 + 0.3*(0.25(4+4) )+ (0.1*2500)/1000;

= 0.00 + 0.6 + 0.25;

= 0.85

**Calculation of Rank Score for "Socialize":**

Since, the word is not matched with Meta data,

We put c =0;

Semantics, s = 0.65;

Visit of links, vl = 0.25;

Google analytics, ga = 0.1;

Rank Score = c*s + vl*(inlinks+outlinks) + (ga*7500)/1000;

= 0*0.65 + 0.3*(0.25(6+3) )+ (0.1*7500)/1000;

= 0.00 + 0.675 + 0.75;

= 1.425

### VIII. RESULT ANALYSIS

Similarly we perform the rank score evaluation for all the other remaining 5 websites. Then we store the rank score values in an array and then we sort the array in ascending order. After that we return the value array [9] which returns the rank score of the maximum value. Our algorithm uses the binary search technique to search for the given word in the Meta data of the website. We also give percentages to other factors like Google analytics and visit of links to help a weak website get a better rank. **We use Binary Search which reduces the complexity of the program to O (logn).This is the best case scenario for the complexity of any algorithm.** We observe here that, "**FoodWorld**" has the **highest rank** score since the word in the **Meta data is matched with the text** entered in the text field.

**Hence the order of ranks will be:**

| Rank 1 | FoodWorld |
|--------|-----------|
| Rank 2 | Socialize |
| Rank 3 | StudyJava |
| Rank 4 | Travelogue |
| Rank 5 | TennisPro |

### IX. CONCLUSION AND FUTURE SCOPE

We had come with an idea of improving the current page ranking algorithm by using a new formula designed by us:

**Rank score = c*s + d*vl*(inlinks+outlinks) + (ga*No. of hits)/1000;**

We calculated the values of few websites from a graph we generated with random inlinks and

outlinks.We gave 65% weightage to semantics, 25% weightage to visit of links and 10% weightage to Google analytics (No. of views).Since, our research was mainly based on semantics page ranking we implemented Binary Search algorithm to reduce our complexity to O (logn).We concluded from our graph and nodes that if a word is matched the rank score is given the highest for that website.

## ACKNOWLEDGEMENT

I would like to take this opportunity to thank everyone including our teachers and parents who believed in us such that we could complete our research work on time and get ourselves ready for a publication. I would like to thank the SSRG journals for publishing our journal and IJCTT for reviewing the same.

## REFERENCES

[1] S. Brin, and Page L., "*The Anatomy of a Large Scale Hypertextual Web Search Engine*", Computer Network and ISDN Systems, Vol. 30, Issue 1-7, pp. 107-117, 1998.
[2] Wenpu Xing and Ghorbani Ali, "*Weighted PageRank Algorithm*", Proceedings of the Second Annual Conference on Communication Networks and Services Research (CNSR '04), IEEE, 2004
[3] Gyanendra Kumar, Neelam Duahn, and Sharma A. K., "*Page Ranking Based on Number of Visits of Web Pages*", International Conference on Computer & Communication Technology (ICCCT)-2011, 978-1-4577-1385-9.
[4] Neelam Tyagi, Simple Sharma, "*Weighted Page Rank Algorithm Based on Number of Visits of Links of Web Page*", International Journal of Soft Computing and Engineering (IJSCE) ISSN: 2231-2307, Volume-2, Issue-3, July 2012.
[5] Robin Nixon, "*Learning PHP, MySQL and Javascript*", Fourth Edition December 2014, O'Reilly.
[6] Narsingh, Deo, "*Graph Theory with Applications to Engineering and Computer Science*" Paperback, 1979.
[7] J. Kleinberg, "*Hubs, Authorities and Communities*", ACM Computing Surveys, 31(4), 1999.
[8] R. Kosala, and H. Blockeel, "*Web Mining Research: A Survey*", SIGKDD Explorations, Newsletter of theACMSpecial Interest Group on Knowledge Discovery and Data Mining Vol. 2, No. 1 pp 1-15, 2000.